\documentstyle[prl,aps,psfig]{revtex}
\begin{document}
\draft
\title{Quantum Hall Spherical Systems: the Filling Fraction}
\author{P. Sitko$\sp{1,2}$,
  J. J. Quinn$\sp{1,3}$, and D. C. Marinescu$\sp{3}$}
\address{
$\sp{1}$University of Tennessee, Knoxville, Tennessee 37996, USA
}
\address{
$\sp{2}$Institute of Physics, Technical University of Wroc{\l}aw, 
Wyb. Wyspia\'nskiego 27, 50-370 Wroc{\l}aw, Poland
}
\address{
$\sp{3}$Oak Ridge National Laboratory, Oak Ridge, Tennessee, 37831, USA
}
\date{\today}
\maketitle

\begin{abstract}
Within the newly formulated composite fermion hierarchy the filling
fraction of a spherical quantum Hall system is obtained when it can be 
expressed as an odd or even denominator fraction. A plot  of 
$\nu\frac{2S}{N-1}$ as a function of 
$2S$ for a constant number of particles
(up to $N=10001$) exhibits structure
of the fractional quantum Hall effect. It is confirmed that $\nu_e +\nu_h=1$
for all particle-hole conjugate systems, except systems
with $N_e =N_h$, and $N_e=N_h \pm 1$. 
\end{abstract}

\narrowtext
During last 15 years systems of the quantum Hall effect
were intensively studied \cite{Prange}, 
both for experimentally related situations
(Laughlin droplet \cite{Laughlin}) and for purely theoretical
spherical systems \cite{Haldane}.
The main problem of any quantum Hall related  studies is the definition
of the filling fraction.
For infinite systems the filling fraction is defined as 
$\nu^{-1}=\frac{B}{\rho}(\frac{hc}{e})^{-1}$, i. e. , the number of the flux 
quanta per electron.
For finite systems, however, this definition can no longer be used.
The only way of assigning the noninteger filling fraction
to the system with given  number of particles
and strength of the magnetic field is either intuitive 
\cite{Laughlin,Haldane} or
 based on the use of the composite fermion transformation
\cite{Jain}.
We extent the formulation of the Jain filling fractions to all odd denominator
hierarchy fractions.
We make use of newly formulated \cite{Sitko2}
composite fermion hierarchy.
The further extension is made by generalizing the composite fermion hierarchy
to all even denominator fractions by introducing the idea
of ``half-filled'' \cite{Rezayi} states of quasiparticles.
In such a way we can obtain the filling fraction for almost all
pairs of $N$ (number of particles) and $2S$ (the strength of the magnetic
monopole for the sphere).

We define the filling fraction in the following way.
First, perform the CF transformation (if $\nu <1$) changing
the value of $2p$ (the strength of the Chern-Simons field) to get
at least one filled effective Landau shell. Hence,
\begin{equation}
\frac{1}{\nu}=2p+\frac{\alpha}{n+\nu_{QE}}\; ,
\end{equation}
where $\alpha$ is the sign of the effective field (with respect
to the real magnetic field), $n$ is the number of filled effective
Landau shells, and $\nu_{QE}$ is the filling fraction for quasielectrons
partially occupying the ``$n+1$'' effective shell.
The process can be repeated on $\nu_{QE}$ until at the $m$-th step
$\nu_{QE}^m=0$ (the hierarchy odd denominator fraction)
or $\alpha_{QE}^m=0$ which corresponds to the ``half-filled''
($\nu_{QE}^m=\frac{1}{2p_{QE}^m}$) quasielectron case (even denominator 
fractions).
Hence, we get all fractions, except the cases when one quasielectron
is left (no fraction can be assigned to one particle system).

In Fig. 1 we plot the values of 
\begin{equation}
\label{function}
\nu\frac{2S}{N-1}
\end{equation}
for eight electrons for values of the filling starting at $\nu=1$
and going  to $\nu=1/5$.
The function (\ref{function}) has the value
of unity when $\nu=\frac{N-1}{2S}$,
which occurs for the Laughlin states ($1$, $1/3$, $1/5$)
and for the ``half-filled'' ($1/2$, $1/4$) states.
For other fractions the function (\ref{function})
varies from unity, but an abrupt change
is clearly visible at $\nu$ close to  $1/2$ and $1/4$.
Such ``discontinuity''  can be explained already with
introduction of Jain states.
The integer filling (for real $2S$ or effective $2S^*$ field)
is obtained when
\begin{equation}
N=n^2 +n(2S)\; ,
\end{equation}
and $N=n^2 +n(2S^* )$ for Jain states.
Since $2S^* =2S-2p(N-1)$ we get
\begin{equation}
\frac{2S}{N-1}=\frac{2pn+1}{n}-\frac{n^2 -1}{n(N-1)}, \;{\rm for}\; 2S^* >0\; ,
\end{equation} 
and
\begin{equation}
\frac{2S}{N-1}=\frac{2pn-1}{n}+\frac{n^2 -1}{n(N-1)}, \;{\rm for}\; 2S^* <0\; .
\end{equation}
Hence, approaching the $1/2p$ states  from both sides we find the corrections
to the filling fractions to be of opposite sign,
with maximum correction for maximum $n$.

We plot similar results for $N=101$ (Fig. 2).
The discontinuity at $1/2p$ states is not only
confirmed by the Jain states but also by other hierarchy states.
Additionally, similar discontinuities can be seen at each 
$1/2p_{QE}$ state of quasielectrons (even denominator fractions).
In the range of $300<2S<500$ the curve clearly repeat the
results for $100<2S<300$, because in the formulation of the fraction
we change only the value of $2p$.
We confirm this by plotting the results for $N=10001$ within the range
$10000<2S<30000$ (Fig. 3), and within the range $30000<2S<50000$ 
(Fig. 4) with adjusted scale for the function (\ref{function}).
Such repeating structure can be also seen when looking at quasielectron
filling, i. e. , $\nu=4/5$ ($\nu_{QE}=1/3$) down to $\nu=2/3$ ($\nu_{QE}=1$).
We plot the results for $N=10001$ and the range $12500<2S<15000$
in Fig. 5. 
The curves are not exactly the same, however, due to the fact that
the number of quasielectrons changes with $2S$, in contrast
to all other figures where $N=const$.

In order to see the structure for $\nu>1$
we plot the results for $N=101$ for $0<2S<500$ in Fig. 6.
A better view is obtained when $N=10001$ and $2000<2S<10000$ in Fig. 7.
The discontinuities again come from analogous  ``half-filled'' states
at higher Landau levels.
The integer fillings are seen as a little jumps at the curve.
Such jumps are also seen in Fig. 2 for Laughlin
and in Fig 3, 4 for Jain states.
In fact, all odd denominator hierarchy states lead to such
jumps (each of them can be seen if the number of particles
is large enough)
and all even denominator fractions give discontinuities when resolution of
the curves (the scale and the number of particles) increases.
Thus, for infinite number of particles
the curve exhibit fractal structure.

In order to confirm our method of defining the filling
fraction we calculate the sum of $\nu_e +\nu_h$ for particle-hole
conjugate states ($\nu_e <1$).
The sum is always one, as expected, except the three cases
when $N_e=N_h$, and $N_e=N_h \pm 1$,
which correspond to difficulty in determing
the filling fraction of $1/2$.
The filling $\nu_e=1/2$ is obtained when $N_e=N_h+1$,
hence, the filling fraction $\nu_h$ for $N_h$ is necessarily less than $1/2$.
Similar problem is for $N_e=N_h$ when the CF hierarchy fraction
is less than $1/2$.
It is worth  noting, however,
that no problems occur at other 
fillings $1/2p$ ($1/4$, $1/6$, ...) which
represent exactly the same problem in terms of composite fermions.


This work was supported in part by Oak Ridge National Laboratory, managed by
Lockheed Martin Energy Research Corp. for the US Department of Energy under
contract No. DE-AC05-96OR22464.
P.S. acknowledges  support by Committee for Scientific Research, Poland,
 grant PB 674/P03/96/10.

\newpage
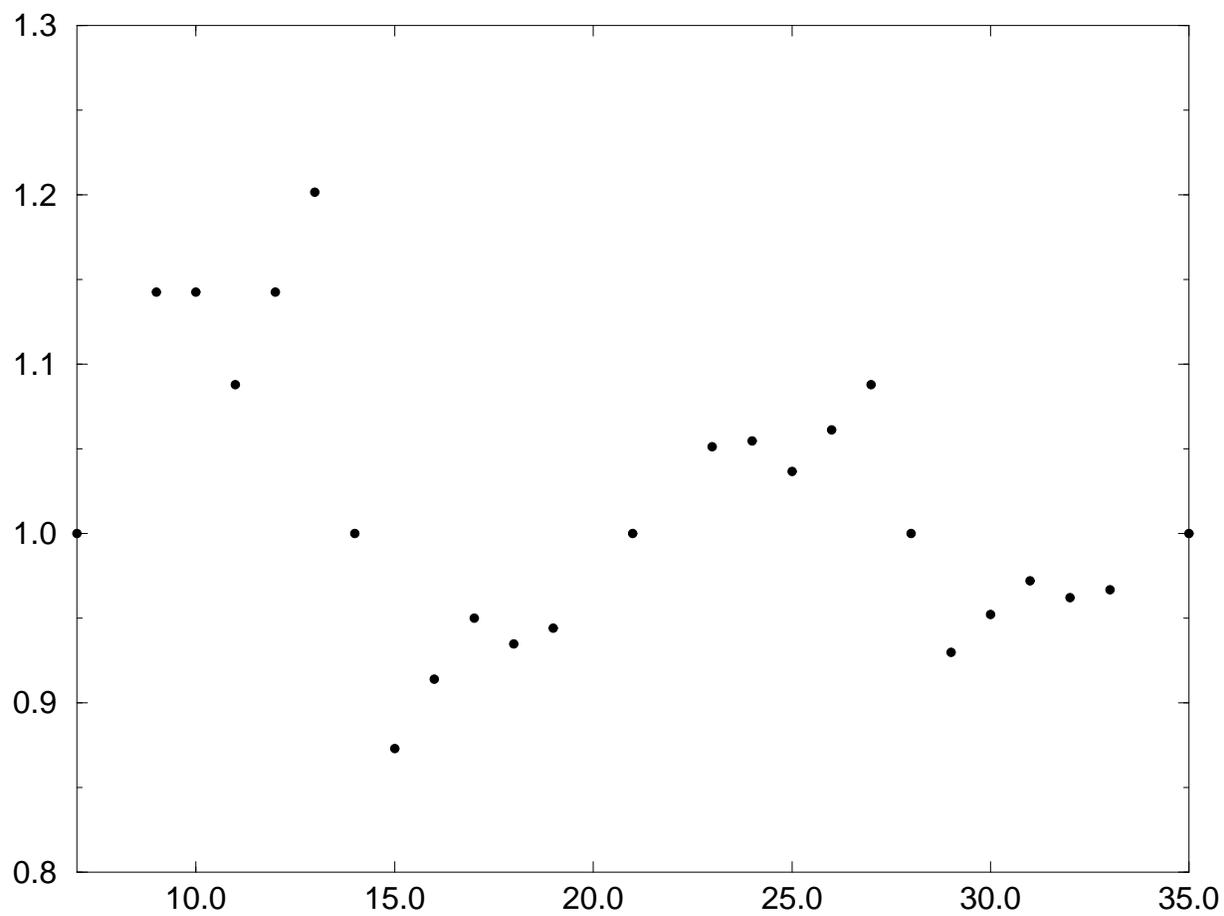
\begin{figure}
\caption{The values of $\nu\frac{2S}{N-1}$ for $N=8$ within
range of $1\ge\nu\ge 1/5$.} 
\end{figure}
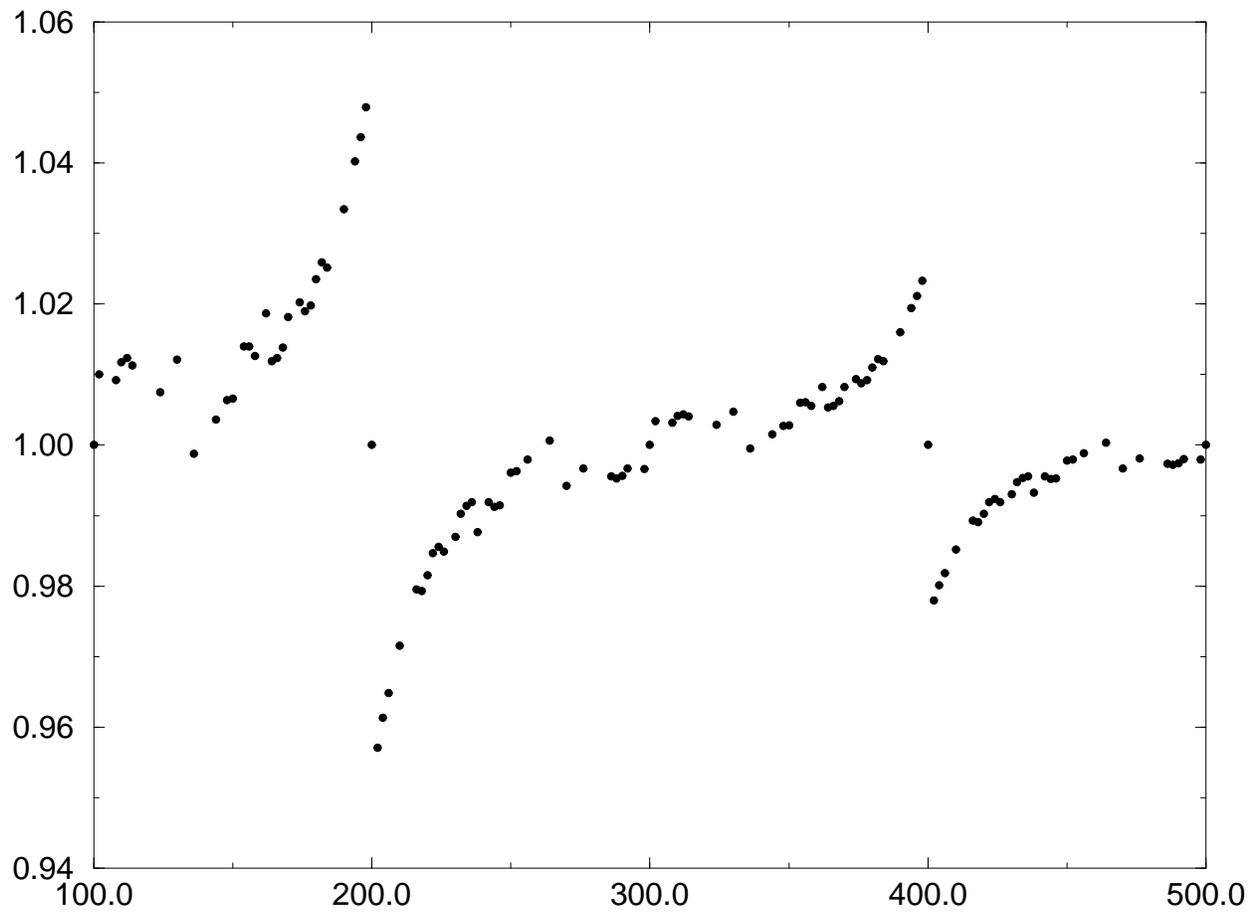
\begin{figure}
\caption{The same as Fig. 1 for $N=101$.}
\end{figure}
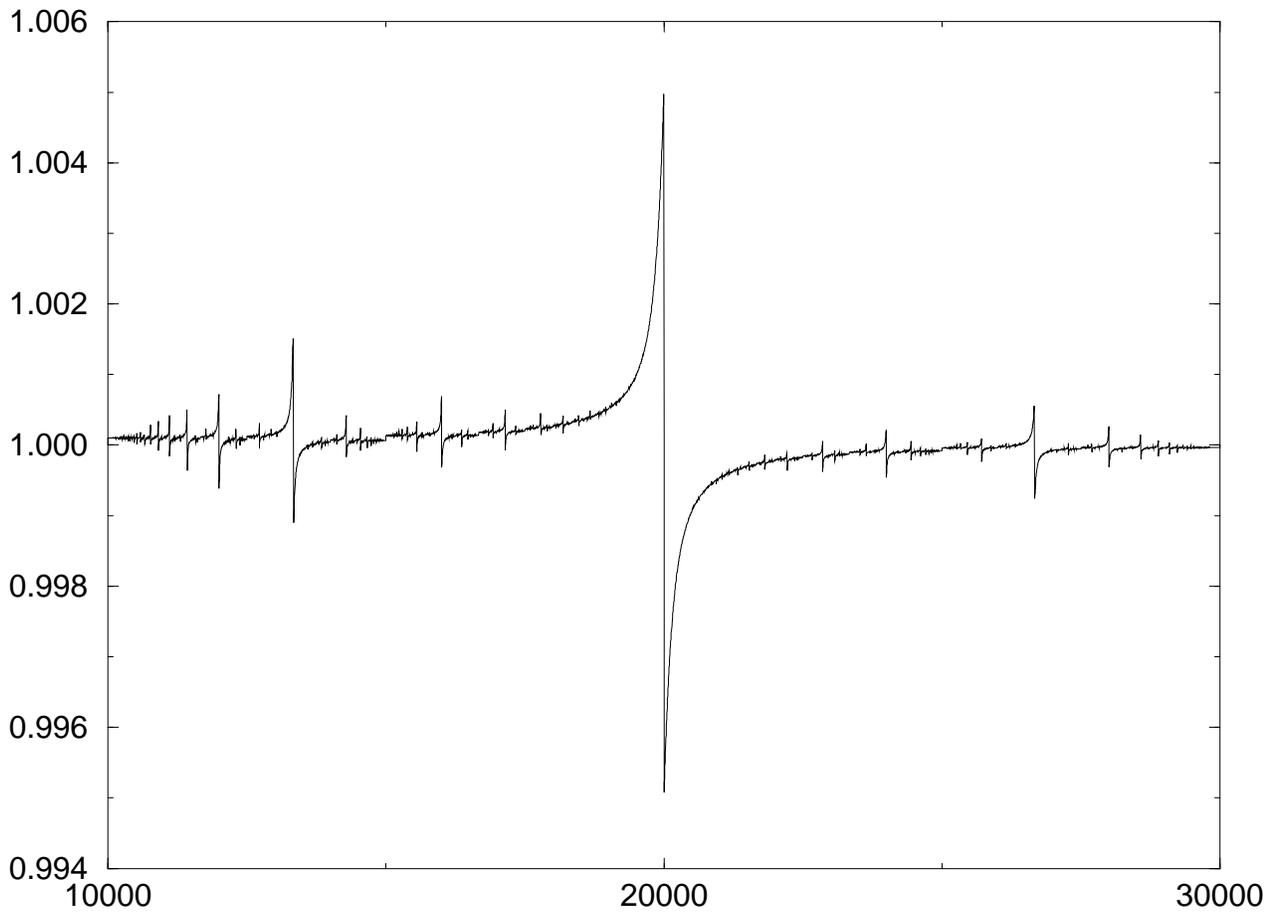
\begin{figure}
\caption{The values of $\nu\frac{2S}{N-1}$ for $N=10001$ within
range of $1\ge\nu\ge 1/3$.}
\end{figure}
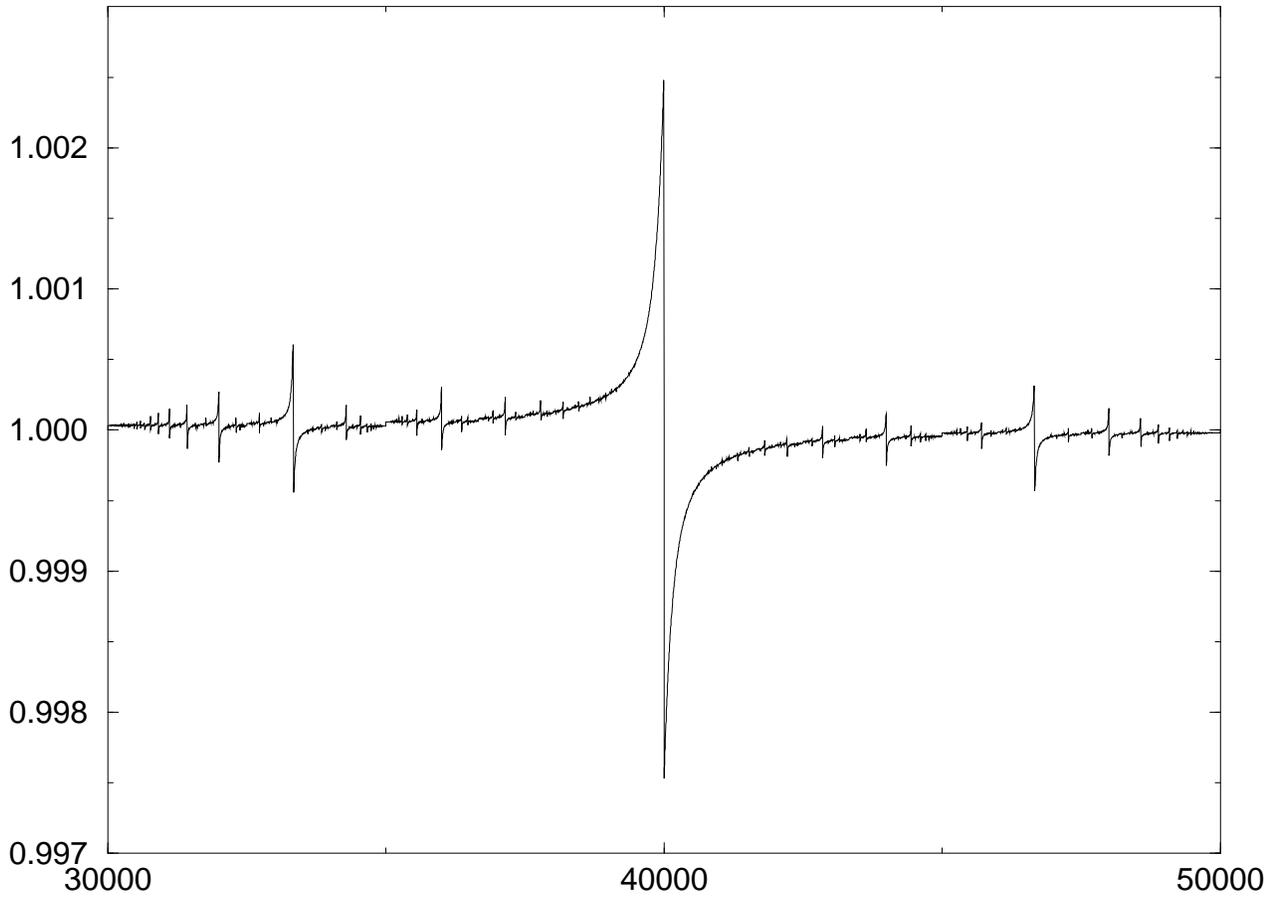
\begin{figure}
\caption{The same as Fig. 3 for $1/3 \ge\nu\ge 1/5$.}
\end{figure}
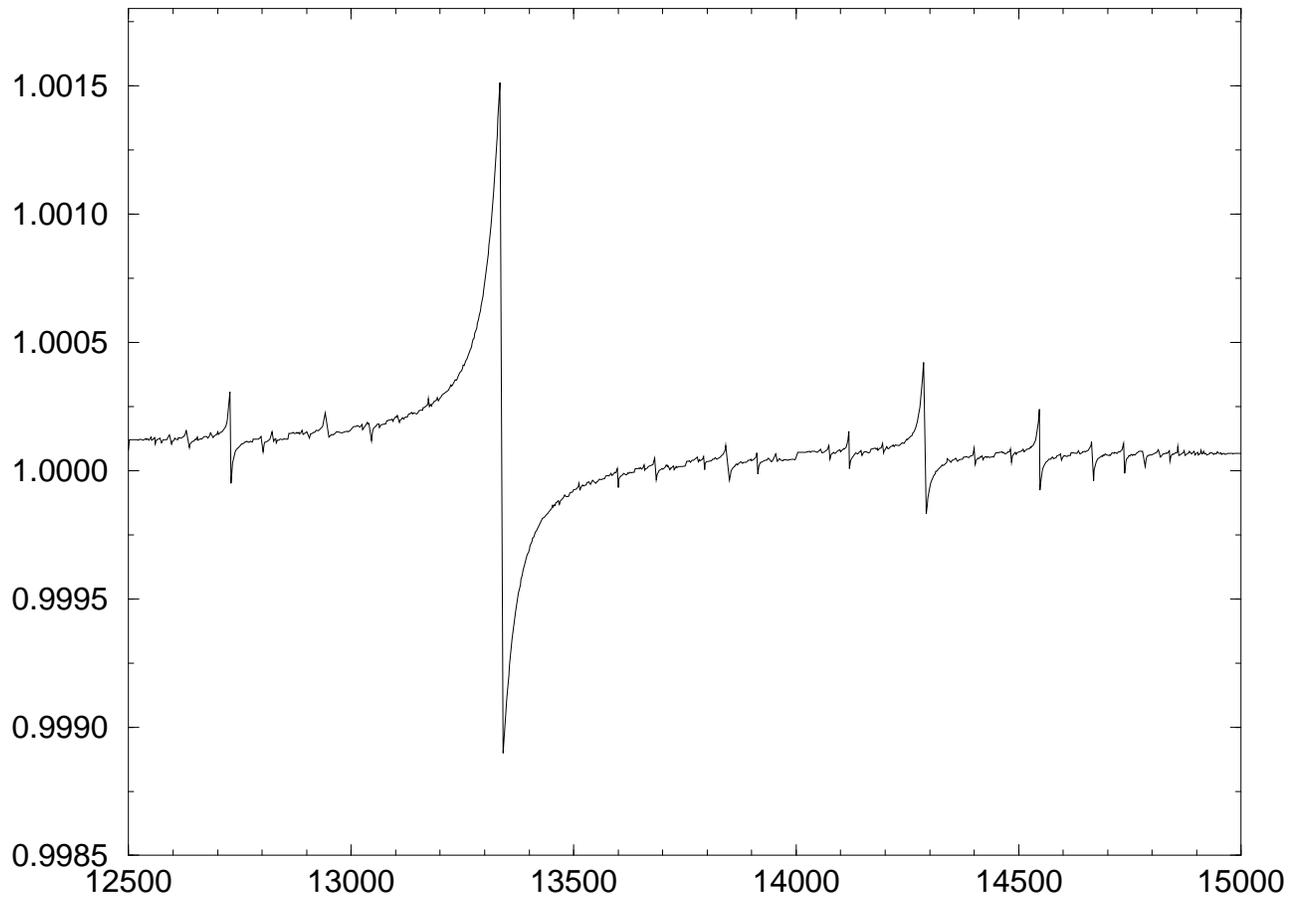
\begin{figure}
\caption{The spectrum of the filling fraction (multiplied by $\frac{2S}{N-1}$)
for $N=10001$ and $12500 \ge 2S\ge 15000$. This represent the systems
of quasielectrons within range of $1\ge \nu_{QE}\ge 1/3$ (note that $\nu_{QE}$
increases with $2S$).}
\end{figure}
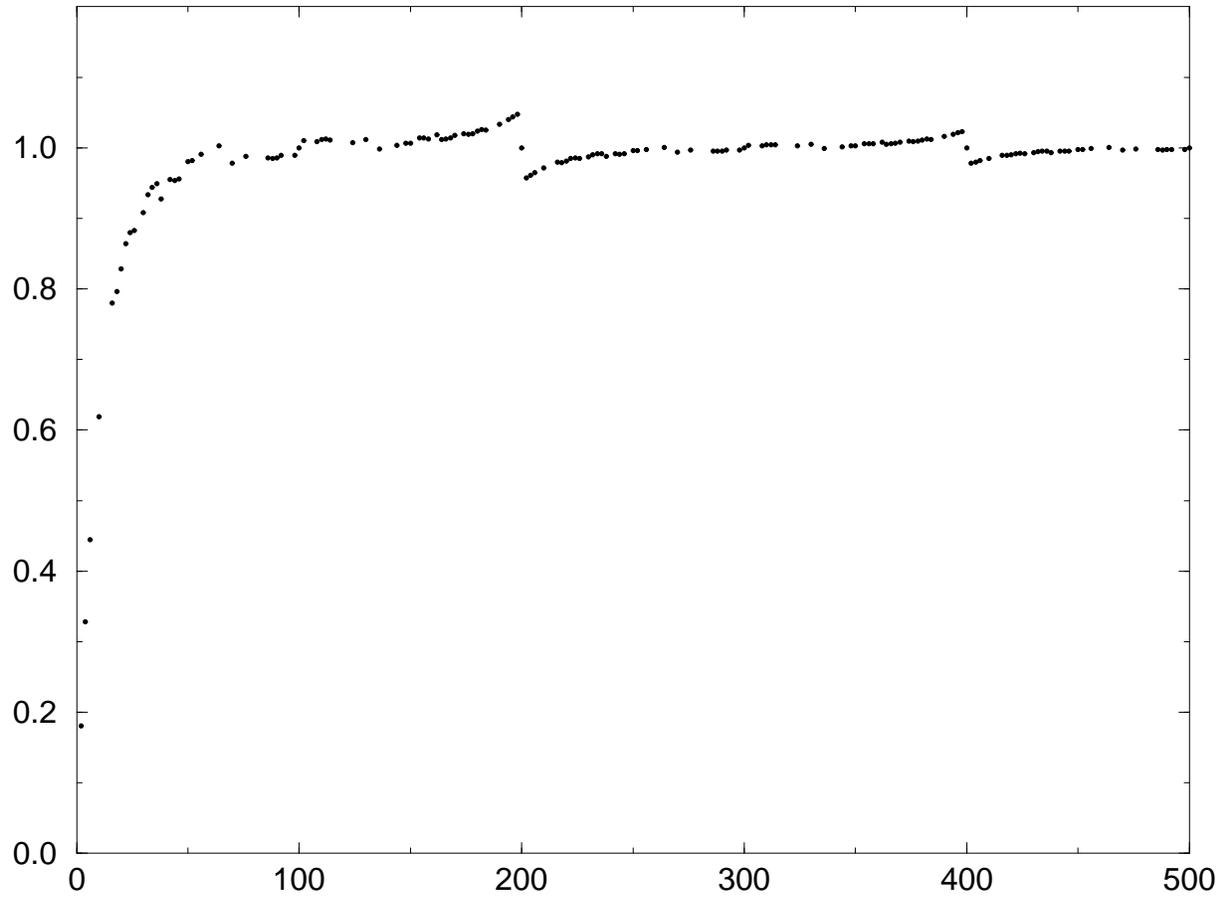
\begin{figure}
\caption{The whole spectrum $0\le 2S\le 500$ for $N=101$.}
\end{figure}
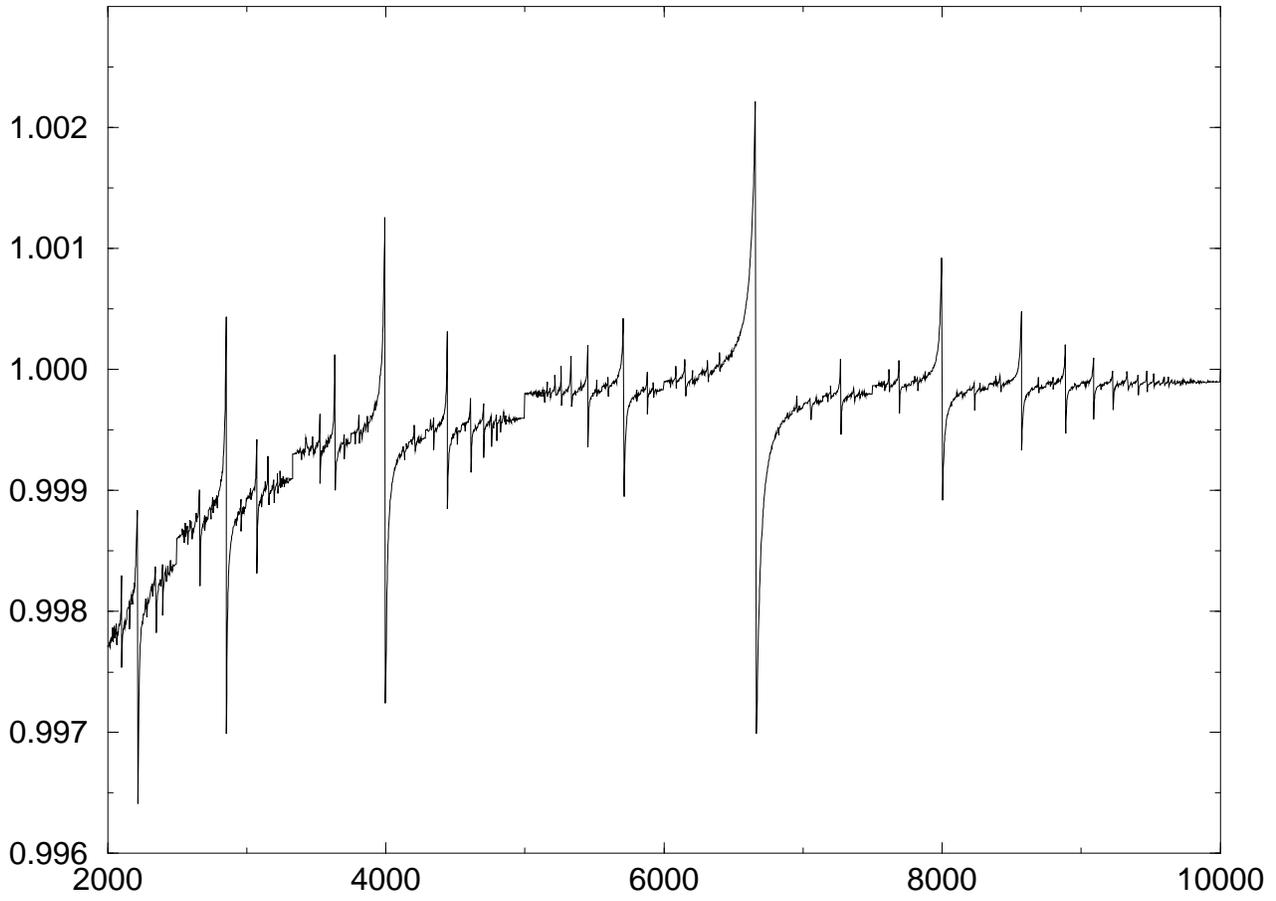
\begin{figure}
\caption{The range of integer filling for $N=10001$, $2000\le 2S\le 10000$.}
\end{figure}

\end{document}